\documentclass[onecolumn,secnumarabic,amsmath,amssymb,balancelastpage,nofootinbib]{article}

\usepackage[e]{esvect}
\usepackage{color}         
\usepackage{graphics}      
\usepackage{graphicx}      
\usepackage{epsf}          
\usepackage{bm}            

\usepackage{cite}
\usepackage{amssymb}
\usepackage{amsmath, esint}
\usepackage{mathrsfs}
\usepackage{framed}
\usepackage{bigints} 
\usepackage{enumitem}
\usepackage{pifont}
\usepackage[capbesideposition={left,center},facing=yes,capbesidewidth=8cm,capbesidesep=quad]{floatrow} 
\usepackage{setspace} 

\usepackage[none]{hyphenat} 

\usepackage[colorlinks=true]{hyperref}  

\usepackage{feynmf}

\definecolor{darkred}{rgb}{0.6,0,0}
\definecolor{darkgreen}{rgb}{0,0.5,0}
\definecolor{darkblue}{rgb}{0,0,0.6}
\hypersetup{ colorlinks,
linkcolor=darkblue,
filecolor=darkgreen,
urlcolor=darkgreen,
citecolor=darkred }

\begin{document}

\sloppy 

\bibliographystyle{nar}

\newlength{\bibitemsep}\setlength{\bibitemsep}{.2\baselineskip plus .05\baselineskip minus .05\baselineskip}
\newlength{\bibparskip}\setlength{\bibparskip}{0pt}
\let\oldthebibliography\thebibliography
\renewcommand\thebibliography[1]{%
  \oldthebibliography{#1}%
  \setlength{\parskip}{\bibitemsep}%
  \setlength{\itemsep}{\bibparskip}%
}


\title{\vspace*{-35 pt}\huge{Eliminating Electron Self-Repulsion}}
\author{Charles T. Sebens\\Division of the Humanities and Social Sciences\\California Institute of Technology}
\date{arXiv v2 - May 4, 2023}

\maketitle
\vspace*{-20 pt}
\begin{abstract}
Problems of self-interaction arise in both classical and quantum field theories.  To understand how such problems are to be addressed in a quantum theory of the Dirac and electromagnetic fields (quantum electrodynamics), we can start by analyzing a classical theory of these fields.  In such a classical field theory, the electron has a spread-out distribution of charge that avoids some of the problems of self-interaction facing point charge models.  However, there remains the problem that the electron will experience self-repulsion.  This self-repulsion cannot be eliminated within classical field theory without also losing Coulomb interactions between distinct particles.  But, electron self-repulsion can be eliminated from quantum electrodynamics in the Coulomb gauge by fully normal-ordering the Coulomb term in the Hamiltonian.  After normal-ordering, the Coulomb term contains pieces describing attraction and repulsion between distinct particles and also pieces describing particle creation and annihilation, but no pieces describing self-repulsion.
\end{abstract}

\tableofcontents
\newpage

\section{Introduction}

There are pervasive problems of self-interaction in both classical and quantum electrodynamics.  In classical electrodynamics, the severity of these problems depends on whether matter is modeled as point charges or as continuous charge distributions.  The latter option dodges the most serious problems of self-interaction, avoiding infinite self-energy and yielding well-defined dynamics that includes radiation reaction.  However, the continuous charge distributions will experience self-repulsion.

If the electron is modeled classically as a cloud of negative charge, the different parts of that cloud would be expected to repel one another.  This self-repulsion would increase the energy of such an electron cloud and would cause the electron to rapidly explode (if there is no force countering the self-repulsion\footnote{Early models of the electron posited forces holding the electron together, now known as ``Poincar\'e stresses'' \cite[ch.\ 28]{feynman2}; \cite{rohrlich1973, rohrlich, pearle1982, schwinger1983electromagnetic}; \cite[ch.\ 16]{jackson}; \cite[sec.\ 5]{griffithsletter}.  However, no such forces appear in quantum electrodynamics or the standard model.  Thus, for our purposes here in understanding the relation between classical and quantum field theories, we need not consider such forces.}).  We do not observe such self-repulsion.  As is apparent in quantum chemistry, the ground state energy of an atom or molecule does not include a contribution from electron self-repulsion.\footnote{The Hartree-Fock method for approximating the ground state energy of an atom or molecule includes contributions from Coulomb repulsion between distinct electrons but explicitly excludes self-repulsion, and a more recent method for calculating ground state energies, called ``density functional theory,'' requires a self-interaction correction that corrects for the initial inclusion of electron self-repulsion \cite[pg.\ 436]{blinder1965}; \cite{perdew1981}; \cite[sec.\ 8.3]{parryang}; \cite[pg.\ 559]{levineQC}; \cite{electronchargedensity}.}

Quantum electrodynamics can be arrived at by quantizing a classical theory of interacting Dirac and electromagnetic fields, where the Dirac field describes charged matter (electrons and positrons) as having a continuous distribution of charge.  Although electron self-repulsion is present in this classical starting point, it should be absent in the quantum field theory that we get upon field quantization: quantum electrodynamics.\footnote{Barut and his collaborators have studied such a theory of interacting classical interacting Dirac and electromagnetic fields, considering it to be an alternative to standard quantum electrodynamics.  They have shown that a number of important phenomena can be explained without quantum physics \cite{barut1989, barut1991}.  These authors retain self-interaction effects, including self-repulsion.  In the context of a discussion of the hydrogen atom, Barut \cite[pg.\ 39]{barut1989} explains why we do not observe electron self-repulsion (even though he thinks that it does in fact occur) as follows: ``in a static situation, the interaction potential between electron and the proton is just $Ze/r$ [where $Z=1$ and $r$ is the distance from the proton], hence we expect that the static self-field of the electron should have no effect - it is already taken into account by the physical charge and mass of the electron.''  I do not see how one could explain the absence of any observed self-repulsion by simply redefining the mass and/or charge of the electron.  The classical Coulomb energy associated with electron self-repulsion \eqref{classicalCoulombenergy} would vary depending on how compactly the electron's charge is spread.}  The question to be tackled here is how self-repulsion is eliminated in quantum field theory when it is present in classical field theory.  Answering this question is important to better understanding the transition from classical to quantum theories of the Dirac and electromagnetic fields, and thus to better understanding quantum field theory.  The classical theory of interacting Dirac and electromagnetic fields is rarely studied, but it is worthy of our attention because of its relation to quantum electrodynamics \cite{fields}.

In section \ref{SIsection}, I briefly discuss the problems of self-interaction that arise for point charges and charge distributions in classical electromagnetism.  In section \ref{CFTsection}, I review the classical theory of the Dirac field interacting with the electromagnetic field, explaining how self-repulsion arises in that context.  Then, in section \ref{QFTsection}, I move to quantum field theory via field quantization in the Coulomb gauge and argue that it is the full normal-ordering of creation and annihilation operators in the Coulomb interaction term of the Hamiltonian that eliminates electron self-repulsion.  The full normal-ordering of the Coulomb term (not merely normal-ordering each instance of the charge density operator) is rarely mentioned and has not been recognized as removing electron self-repulsion.  To understand the role of the normal-ordered Coulomb term in the Hamiltonian, in section \ref{QFTsection} we first imagine that it is the only interaction term in a simplified quantum theory of the Dirac field (that could be called ``quantum electrostatics'') and then move to full quantum electrodynamics.

Before diving in, let me clarify the kind of explanation of the absence of electron self-repulsion that is being sought in this paper.  Ultimately, the presence or absence of self-repulsion will be a consequence of the dynamical laws of a given theory.  The dynamical laws of classical electrodynamics include self-repulsion.  The dynamical laws of quantum electrodynamics, properly formulated, avoid self-repulsion.  I take the laws posited by a theory to be brute facts about the world that do not themselves require explanation\footnote{See \cite[ch.\ 4]{lange}.} (at least not if that theory is being treated as fundamental physics).  Thus, finding that the laws of quantum electrodynamics avoid self-repulsion is sufficient for explaining the absence of electron self-repulsion in nature.

The goal of this paper is to understand how electron self-repulsion is removed in the shift from classical to quantum field theories, not to resolve all puzzles of self-interaction in quantum electrodynamics.  In textbook discussions of Feynman diagram methods for calculating scattering amplitudes in quantum electrodynamics, the electron is standardly described as having an infinite self-energy that must be rendered finite via some procedure of mass renormalization \cite[sec.\ 15a]{schweber1961}; \cite[ch.\ 8]{bjorkendrell}; \cite[sec.\ 5.3]{greiner2003}; \cite[sec.\ 18.2]{schwartz}.  This is somewhat surprising if you take the starting point for quantum electrodynamics to be a classical theory of interacting Dirac and electromagnetic fields where electron charge is spread out and the electrostatic self-energy is finite.  However, the relation between this infinite quantum self-energy and the classical electromagnetic self-energy of a point charge or charge distribution is not entirely clear.  Schweber \cite[pg.\ 514]{schweber1961}, for example, writes that ``the self-energy problem is of a rather different nature in quantum theory from that found in classical theory'' and remarks that it ``seems unlikely that a solution of the classical self-energy problem is either necessary or sufficient for the solution of the quantum mechanical one.''  I believe that more work needs to be done to better understand the connections (or lack thereof) between the classical and quantum self-energy problems.  This project is intended to be just one step towards that deeper understanding.

\section{Self-Interaction and Self-Repulsion}\label{SIsection}

\begin{figure}[htb]
\center{\includegraphics[width=6 cm]{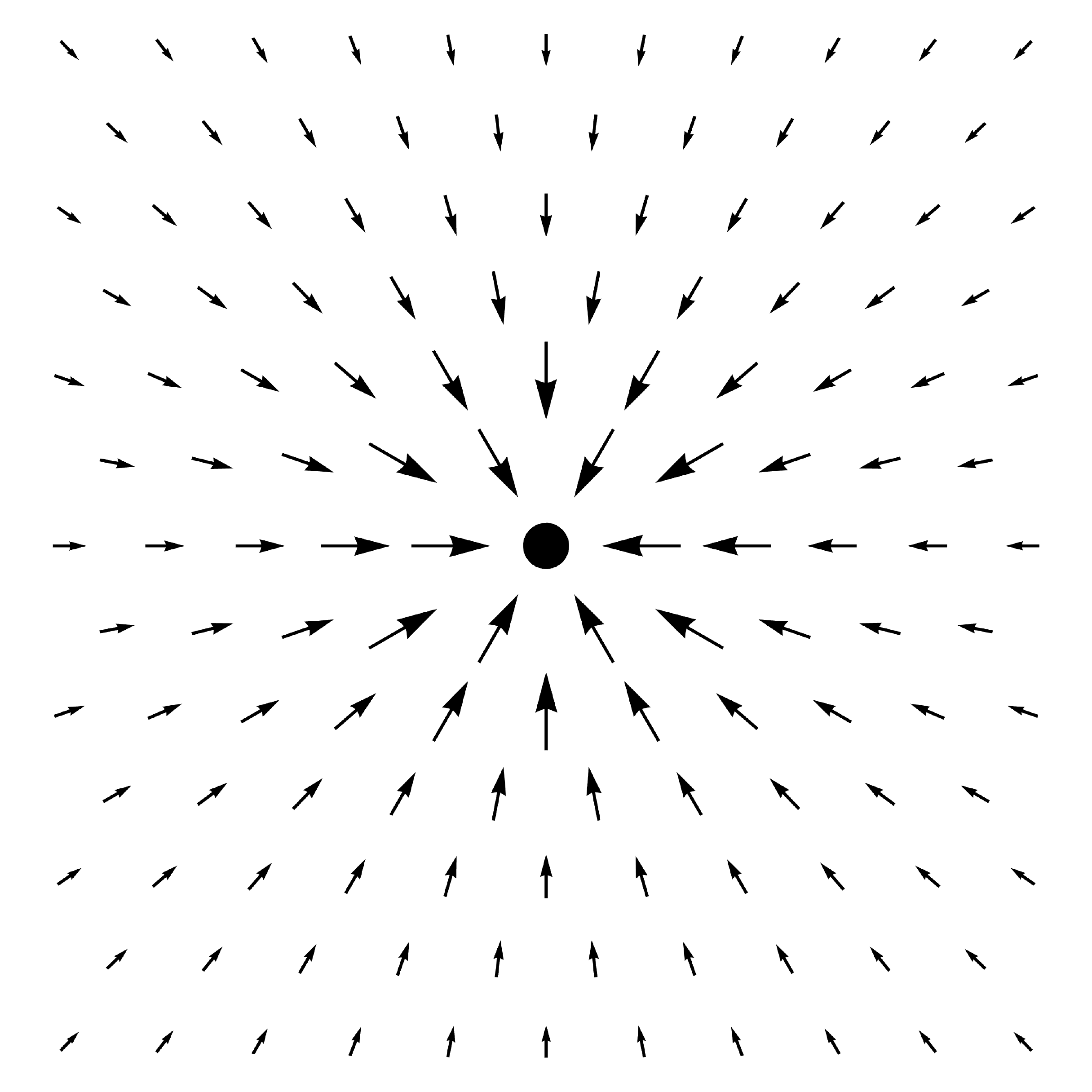}\hspace*{24 pt}\includegraphics[width=6 cm]{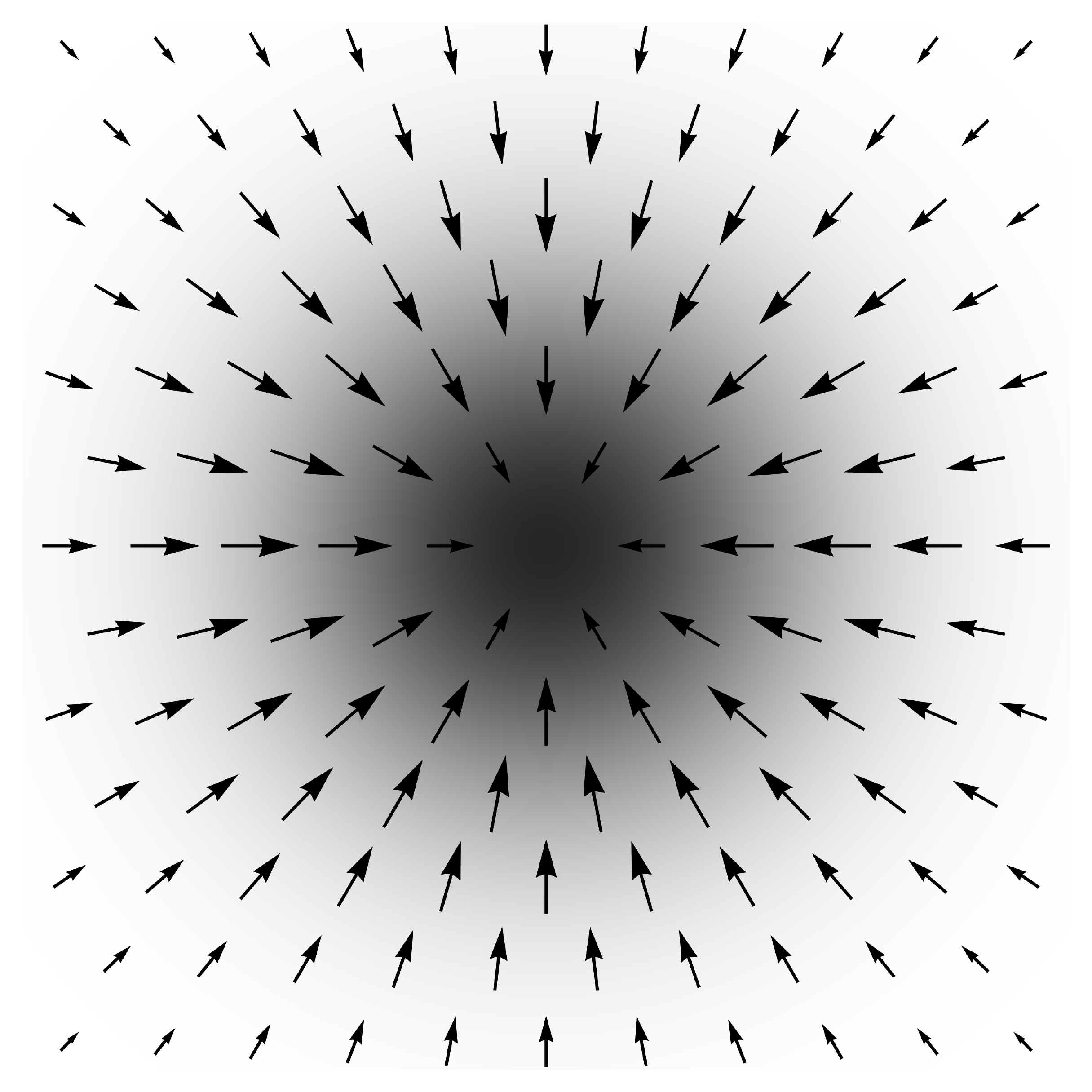}}
\caption{The left image depicts the electric field of a point negative charge at rest.  The image on the right depicts the electric field of a static Gaussian distribution of negative charge (shown as a gray cloud).}
  \label{selffield}
\end{figure}

In classical electromagnetism, one can take the charged matter interacting with the electromagnetic field to be either point charges or charge distributions \cite{maudlin2018}.  Point charges come with a variety of problems of self-interaction.  The electric field of a point charge increases in strength as one approaches the charge, becoming infinite in strength and indeterminate in direction at the location of the charge (figure \ref{selffield}).  The density of energy in the charge's electromagnetic field is $\frac{E^2}{8 \pi} + \frac{B^2}{8 \pi}$, which yields an infinite amount of ``self-energy'' when integrated over any finite volume around the charge.  A consequence of the electric field not being well-defined at the location of the point charge is that the standard Lorentz force, $\vec{F}=q \vec{E} + \frac{q}{c}\vec{v}\times\vec{B}$, does not yield a well-defined force on the point charge from its own field.  One could attempt to alleviate this problem by positing that a point charge does not react to its own field, but then one would miss an important effect: radiation reaction.  To accelerate a charged body from rest, you must put in additional energy, beyond the energy acquired by the body, to produce the electromagnetic wave that is emitted when the body accelerates (carrying energy away).  If charges only react to one another's fields through the Lorentz force law, energy-bearing electromagnetic waves will be produced without that additional energy having to be put in.  Because deleting self-interaction leads to violations of energy conservation, Frisch \cite{frisch2004, frisch2005, frisch2008, frisch2009} has described classical electromagnetism as an inconsistent theory (see also \cite{belot2007}).  There are a variety of strategies that might be pursued to fix-up classical electromagnetism with point charges \cite[sec.\ 3]{lazarovici2018}, but let us not wade into those waters here.  Our focus will be on charge distributions because we will be viewing quantum field theory as built from a classical field theory where charge is spread-out  (not concentrated at points).  In that context, there is only one problem of self-interaction: self-repulsion.

Moving away from point charges, we can model the electron classically as a spread-out cloud-like charge distribution (figure \ref{selffield}).  The electric field of the electron grows in strength as one approaches from far away, eventually reaching a maximum strength and then becoming weaker as one moves towards the center of the electron.  Integrating the energy density $\frac{E^2}{8 \pi} + \frac{B^2}{8 \pi}$ yields a finite amount of energy in the electron's electromagnetic field.  The Lorentz force density, $\vec{f}=\rho \vec{E} + \frac{\rho}{c}\vec{v}\times\vec{B}$, is always well-defined.\footnote{Given that the electron will be modeled classically in the next section as a lump of energy and charge in the classical Dirac field, one might wonder whether it makes sense to think of the above density of force as acting on a field (the Dirac field).  For defense of the idea that forces act on both the electromagnetic and Dirac fields, see \cite{forcesonfields, spinmeasurement}.}  If such an electron charge distribution is accelerated, the electromagnetic wave that is emitted will exert a force on the electron as it exits.  Thus, the electron will experience a radiation reaction force and energy will be conserved.  The problems for point charges presented in the previous paragraph are not problems for such a charge distribution.

Still, there is a problem of self-interaction for a spread-out electron charge distribution.  Looking at figure \ref{selffield}, the electric field at any location within the electron points inward and thus the $\rho \vec{E}$ electric force density points outward.  These outward forces are forces of self-repulsion.  One can perfectly well include forces of self-repulsion without having any ill-defined or infinite forces.  Indeed, it makes sense to do so when modeling macroscopic bodies.  However, we know that electrons do not experience self-repulsion and thus the presence of such forces here is a cause for concern.  We must show that this self-repulsion, though present in classical field theory, is absent in quantum field theory.  In the next section, we will fill in the above classical picture of electrons with spread-out distributions of charge, using the classical Dirac field to represent electrons.  Then, in section \ref{QFTsection} we will move to quantum field theory and see how self-repulsion can be eliminated.

\section{Classical Field Theory}\label{CFTsection}

Textbooks on quantum field theory generally begin their treatment of electrons and positrons with a discussion of the Dirac equation, viewed as giving the time-evolution of a four-component complex-valued\footnote{Although I will not do so here, there are reasons (coming from quantum field theory) to treat the components of the Dirac field $\psi$ at a given location as anticommuting Grassmann numbers instead of complex numbers (see \cite{fields} and references therein).} entity $\psi$.  Dirac originally thought of $\psi$ as a single-electron wave function, in which case the Dirac equation would be part of a relativistic single-particle quantum theory.  Alternatively, $\psi$ can be thought of as a classical field (like the electromagnetic field) and the Dirac equation as part of a classical field theory (playing a similar role to Maxwell's equations).\footnote{These two interpretations of the Dirac equation and their relation to quantum field theory are discussed in \cite{fields}.}  Here we will adopt the latter interpretation and start with a classical theory of the Dirac field.  Allowing this field to interact with the electromagnetic field, we have a classical field theory that can be used to arrive at quantum electrodynamics through field quantization.  This classical field theory is sometimes called ``Maxwell-Dirac theory.''

Before proceeding, let me emphasize that classical Maxwell-Dirac field theory is of interest not because it itself accurately describes the behavior of electrons, but because studying it can shed light on quantum field theory.  The classical and quantum field theories are deeply connected.  It is classical Maxwell-Dirac field theory that yields quantum electrodynamics upon field quantization.  That being said, it is important to note that one cannot ``go the other way'' and derive Maxwell-Dirac field theory in the classical limit as an approximation to quantum electrodynamics \cite[pg.\ 221]{duncan}.

In classical Maxwell-Dirac field theory, the Dirac field $\psi$ evolves in accordance with the Dirac equation (including interactions with the electromagnetic field):
\begin{equation}
i \hbar \frac{\partial \psi}{\partial t} = \big(-i \hbar c\, \gamma^0\vec{\gamma}\cdot\vec{\nabla} + \gamma^0 m c^2\big)\psi+ e\, \gamma^0\vec{\gamma}\psi \cdot \vec{A}-e\,\psi\phi
\ .
\label{diracequation}
\end{equation}
The electromagnetic field evolves by Maxwell's equations, with the charge and current densities of the Dirac field,
\begin{align}
\rho&=-e\psi^\dagger\psi
\label{chargedensity}
\\
\vec{J}&=-ec \psi^\dagger\gamma^0\vec{\gamma}\psi
\label{currentdensity}
\ ,
\end{align}
acting as source terms.\footnote{In \cite{positrons} I propose changing the charge and current densities so that the (free) classical Dirac field describes both negatively charged electrons and positively charged positrons.  For our purposes here, where we are concerned first and foremost with electrons, we can stick with the standard charge and current densities.  However, ultimately it would be better to extend the treatment in \cite{positrons} to include interactions with the electromagnetic field.}  In \eqref{diracequation}, electromagnetic interactions are written in terms of the vector and scalar potentials, which specify the electric and magnetic fields via
\begin{align}
\vec{E}&=-\vec{\nabla}\phi-\frac{1}{c}\frac{\partial\vec{A}}{\partial t}
\label{Efieldfrompotentials}
\\
\vec{B}&=\vec{\nabla} \times \vec{A}
\ .
\label{Bfieldfrompotentials}
\end{align}
All equations in this paper appear in Gaussian cgs units.

The Hamiltonian, giving the total energy for the interacting Dirac and electromagnetic fields, is\footnote{In \cite{potentialenergy} I discuss the interpretation of different terms in this Hamiltonian, arguing that the first two terms give the energy of the electromagnetic field and the remainder gives the energy of the Dirac field (with the final term being a potential energy of the Dirac field).}
\begin{equation}
H=\int \left( \frac{E^2}{8 \pi} + \frac{B^2}{8 \pi} +\psi^\dagger\big(-i \hbar c\, \gamma^0\vec{\gamma}\cdot\vec{\nabla} + \gamma^0 m c^2\big)\psi+ e\, \psi^\dagger\gamma^0\vec{\gamma}\psi \cdot \vec{A}\right)d^3 x
\ .
\label{hamiltonian}
\end{equation}
Adopting the Coulomb gauge\footnote{For presentations of quantum electrodynamics in the Coulomb gauge (including discussion of the Hamiltonian for interacting Dirac and electromagnetic fields), see \cite[sec.\ 15.2 and 17.9]{bjorkendrellfields}; \cite[sec.\ 5.2 and 8.1]{hatfield}; \cite[sec.\ 8.3]{weinbergQFT}; \cite[sec.\ 6.4]{tong}.  To limit the scope of this paper, I will work entirely within the Coulomb gauge and not ask how the points made here might be extended to other gauges.} ($\vec{\nabla}\cdot \vec{A}=0$) and using Gauss's law ($\vec{\nabla}\cdot\vec{E} = 4 \pi \rho$) as well as \eqref{chargedensity}, \eqref{Efieldfrompotentials}, and \eqref{Bfieldfrompotentials}, it is possible to rewrite this Hamiltonian exclusively in terms of $\vec{A}$ and $\psi$,
\begin{align}
H&=\int \left( \frac{\big|\frac{\partial\vec{A}}{\partial t}\big|^2}{8 \pi c^2}
 + \frac{\big(\vec{\nabla} \times \vec{A}\,\big)^2}{8 \pi} +\psi^\dagger\big(-i \hbar c\, \gamma^0\vec{\gamma}\cdot\vec{\nabla} + \gamma^0 m c^2\big)\psi\right.
 \nonumber
 \\
&\qquad\quad\left. + e\, \psi^\dagger\gamma^0\vec{\gamma}\psi \cdot \vec{A}+ \frac{e^2}{2} \int \frac{\psi^{\dagger}(\vec{x})\psi(\vec{x})\psi^{\dagger}(\vec{y})\psi(\vec{y})}{|\vec{x}-\vec{y}\,|} d^3 y\right)d^3 x
\ .
\label{hamiltonian2}
\end{align}
The last term gives the standard classical energy of Coulomb repulsion for the (always negative) charge density \eqref{chargedensity} of the Dirac field,
\begin{equation}
\frac{1}{2} \int \frac{\rho(\vec{x})\rho(\vec{y})}{|\vec{x}-\vec{y}\,|}d^3 x d^3 y
\ .
\label{classicalCoulombenergy}
\end{equation}
It is this term that captures the energy of Coulomb repulsion between electrons and also, inconveniently, the energy of electron self-repulsion.

To see the problem, let us first decompose the Dirac field via the standard plane wave expansion,
\begin{align}
\psi (\vec{x}) &=\overbrace{\frac{1}{(2\pi\hbar)^{3/2}}\int{ \frac{d^3 p}{\sqrt{2 \mathcal{E}(\vec{p}\,)}} \sum_{s=1}^2 \left(b^{s} (\vec{p}\,) \: u^s (\vec{p}\,) \, e^{\frac{i}{\hbar} \vec{p} \cdot \vec{x}}\right)}}^{\mbox{$\psi_{+}(\vec{x})$}}
\nonumber
\\
&\qquad+\underbrace{\frac{1}{(2\pi\hbar)^{3/2}}\int{ \frac{d^3 p}{\sqrt{2 \mathcal{E}(\vec{p}\,)}} \sum_{s=1}^2 \left(d^{s \dagger} (\vec{p}\,) \: v^s (\vec{p}\,) \,  e^{-\frac{i}{\hbar} \vec{p} \cdot \vec{x}}\right)}}_{\mbox{$\psi_{-}(\vec{x})$}}
\ ,
\label{planewaveexpansion}
\end{align}
where $s$ is a spin index, $\mathcal{E}(\vec{p}\,) = \sqrt{m^2 c^4 + |\vec{p}\,|^2 c^2}$, $u^s (\vec{p}\,)$ and $v^s (\vec{p}\,)$ are basis spinors, and $b^s (\vec{p}\,)$ and $d^{s\dagger} (\vec{p}\,)$ are complex coefficients specifying the particular state of the Dirac field.  In the absence of electromagnetic interactions, $\psi_{+}(\vec{x})$ is a sum of positive-frequency (electron) modes with different momenta $\vec{p}$, each evolving by $e^{-\frac{i}{\hbar}\mathcal{E}(\vec{p}\,)t}$ independently of one another and independently of the negative-frequency (positron) modes, $\psi_{-}(\vec{x})$.  Including interactions, the evolution of $\psi_{+}(\vec{x})$ and $\psi_{-}(\vec{x})$ becomes more complicated.

Let us take a state of the classical Dirac field representing a single electron and nothing else to be a state where the positron part of the Dirac field, $\psi_{-}(\vec{x})$, is zero everywhere and the integral of the charge density \eqref{chargedensity}, $-e\psi_+^\dagger(\vec{x})\psi_+(\vec{x})$, over all space is $-e$ (the charge of the electron) \cite{howelectronsspin, smallelectronstates, spinmeasurement}.  For such a state, the Coulomb term in the Hamiltonian \eqref{classicalCoulombenergy} gives the energy of self-repulsion for the electron's cloud of negative charge,
\begin{equation}
\frac{1}{2} \int \frac{\rho(\vec{x})\rho(\vec{y})}{|\vec{x}-\vec{y}\,|}d^3 x d^3 y=\frac{e^2}{2} \int \frac{\psi_+^\dagger(\vec{x})\psi_+(\vec{x})\psi_+^\dagger(\vec{y})\psi_+(\vec{y})}{|\vec{x}-\vec{y}\,|}d^3 x d^3 y
\ .
\label{electronrepulsion}
\end{equation}
For a state of the Dirac field representing two electrons, where $\int -e\psi_+^\dagger(\vec{x})\psi_+(\vec{x}) \, d^3 x=-2e$, the Coulomb term \eqref{classicalCoulombenergy} includes three kinds of energy: the energy of self-repulsion for the first electron, the energy of self-repulsion for the second electron, and the energy of repulsion between the two electrons.  Ideally, we would like to only include the third kind of energy.

\begin{figure}[htb]
\center{\includegraphics[width=12 cm]{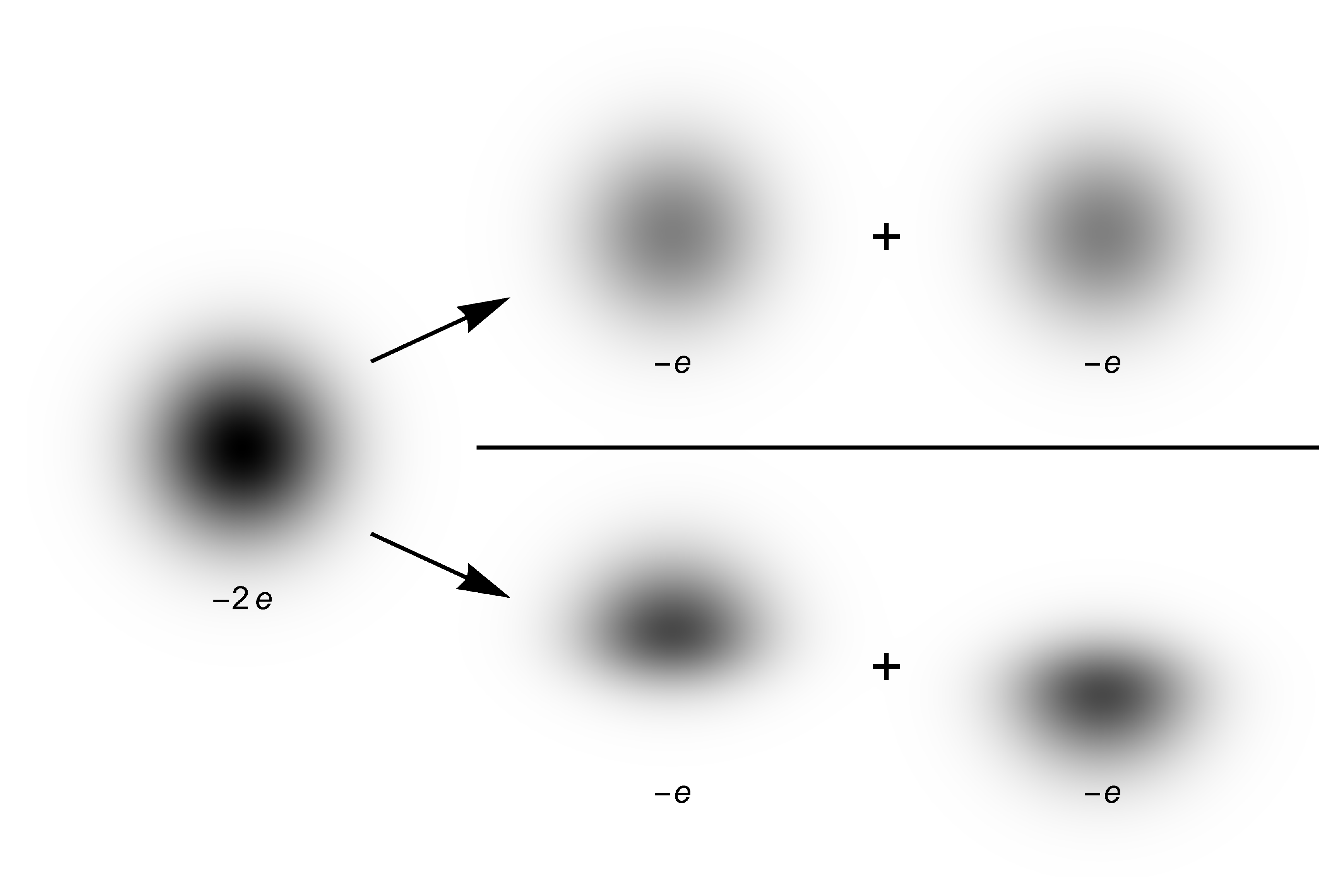}}
\caption{If the classical Dirac field has a spherically symmetric charge density with total charge $-2e$, there are many ways that one could imagine decomposing that charge density into separate contributions from two distinct electrons.  There might be two electrons with the same charge distribution, or, one electron that is responsible for the top half of the total charge distribution and another that is responsible for the bottom half.  The second alternative would have a larger energy of self-repulsion and a smaller energy of repulsion between the electrons (with the total energy of electrostatic repulsion the same as in the first alternative).}
  \label{fieldseparation}
\end{figure}

Within a classical theory of the electromagnetic and Dirac fields, there is no way to include the energy of Coulomb repulsion between electrons while excluding the self-repulsion within each electron's charge distribution.  The problem stems from the lack of separation between electrons in the classical Dirac field.  If we have a state of the Dirac field with total charge $-2e$ representing two electrons, what is the contribution of each electron to the total charge distribution?  There would be multiple ways to divide the total charge density into a density of charge for one electron and a density of charge for the other (figure \ref{fieldseparation}).  This is a problem with classical field theory that we will shortly see resolved in quantum field theory, where it is possible to eliminate self-repulsion from the Hamiltonian.

It is worth noting that self-repulsion can be removed in classical field theory if one shifts from a theory where there is a single Dirac field to a theory where there are many Dirac fields, one for each electron or positron.  One could then modify the Hamiltonian so that the only contributions to the Coulomb energy \eqref{classicalCoulombenergy} come from multiplying the charge densities of distinct Dirac fields.  Let us set this option for removing self-repulsion aside because it is a significant departure from the classical field theory being considered and it is a departure that seems to fit poorly with quantum electrodynamics, where there appears to be only a single (quantum) Dirac field.

\section{Quantum Field Theory}\label{QFTsection}

Moving to quantum field theory allows us to do something that could not be done within classical field theory: including Coulomb attraction and repulsion between distinct electrons while excluding self-repulsion (and retaining radiation reaction).  We will see that self-repulsion can be eliminated by normal-ordering the Coulomb term in the Hamiltonian.  Although I have seen the Coulomb term written with the requisite normal-ordering in one place \cite[eq.\ 8.28]{hatfield}, I have not seen any author explain how this maneuver removes electron self-repulsion.\footnote{Greiner and Reinhardt \cite[sec.\ 8.6]{greiner1996} do not explicitly write down the normal-ordered Coulomb term, but they do seem to think that the entire Hamiltonian (in both spinor and scalar quantum electrodynamics) should be fully normal-ordered.  Greiner and Reinhardt \cite[pg.\ 238]{greiner1996} explain that normal-ordering removes an undesirable kind of self-interaction, writing that ``the prescription of normal-ordering of the interaction operator eliminates the interaction of a particle with itself at the same point $x$.''  However, they do not directly discuss self-repulsion.}

In quantum field theory, the time evolution of a quantum state\footnote{There are a number of competing proposals as to the nature of these quantum states.  Elsewhere, I have argued that states in quantum field theory should be viewed as wave functionals assigning quantum amplitudes to classical field configurations \cite{fields}.} is given by a Schr\"{o}dinger equation of the general form
\begin{equation}
i \hbar \frac{d}{dt} | \Psi (t) \rangle = \widehat{H} | \Psi (t) \rangle
\ .
\label{schrodinger}
\end{equation}
For quantum electrodynamics in the Coulomb gauge, the Hamiltonian is an operator version of the classical Hamiltonian discussed in the previous section \eqref{hamiltonian2}.  The Coulomb term becomes
\begin{equation}
\frac{e^2}{2} \int \frac{:\widehat{\psi}^{\dagger}(\vec{x})\widehat{\psi}(\vec{x})\widehat{\psi}^{\dagger}(\vec{y})\widehat{\psi}(\vec{y}):}{|\vec{x}-\vec{y}\,|}d^3 x\,  d^3 y
\ ,
\label{coulombnormal}
\end{equation}
where the colons indicate normal-ordering (rearranging so that creation operators appear to the left of annihilation operators, including a minus sign whenever a creation operator for an electron or positron is moved past an annihilation operator).  Often, the normal-ordering of the Coulomb term is not explicitly mentioned when the Hamiltonian for quantum electrodynamics is presented.\footnote{See \cite[eq.\ 8.3.12]{weinbergQFT}; \cite[sec.\ 6.4]{tong}; \cite{kay2020}.}  Bjorken and Drell\cite[eq.\ 15.28 and 17.60]{bjorkendrellfields} give what I believe to be an incorrect\footnote{When Bjorken and Drell \cite[sec.\ 17.9]{bjorkendrellfields} use the Coulomb term in deriving the Feynman rules for quantum electrodynamics, they only consider its contribution to diagrams involving interactions between distinct particles.  Thus, although they do not explicitly normal order the entire term, they treat it as if it was normal-ordered.} version of the Coulomb term, separately normal-ordering each appearance of the charge density operator:
\begin{equation}
\frac{e^2}{2} \int \frac{:\widehat{\psi}^{\dagger}(\vec{x})\widehat{\psi}(\vec{x}):\ :\widehat{\psi}^{\dagger}(\vec{y})\widehat{\psi}(\vec{y}):}{|\vec{x}-\vec{y}\,|}d^3 x\,  d^3 y = \frac{1}{2} \int \frac{\widehat{\rho}(\vec{x})\widehat{\rho}(\vec{y})}{|\vec{x}-\vec{y}\,|}d^3 x\,  d^3 y
\ ,
\label{coulombnormalbad}
\end{equation}
where the charge density operator is
\begin{equation}
\widehat{\rho}(\vec{x})=-e:\widehat{\psi}^{\dagger}(\vec{x})\widehat{\psi}(\vec{x}):
\ .
\end{equation}
This partially normal-ordered Coulomb term \eqref{coulombnormalbad} is the operator version of the classical energy of electric repulsion in \eqref{classicalCoulombenergy}. Like that equation, it includes a form of electron self-interaction.  We will see shortly that it is the full normal-ordering in \eqref{coulombnormal} that allows us to avoid this self-interaction.

Before analyzing the role that the Coulomb term \eqref{coulombnormal} plays in quantum electrodynamics, let us first imagine that this term describes the only way in which electrons and positrons interact and ask how the free dynamics are altered.  Put another way, we can start with a free quantum theory of the Dirac field and then see how the dynamics change if the Coulomb term is added.  When this term is added, we get a quantum theory of a single self-interacting field that might be called ``quantum electrostatics,'' because the only interaction that is included is the Coulomb interaction.  (Although the title ``quantum electrostatics'' is not in wide use, Kay \cite{kay2020} has used it to refer to a different theory: a quantum theory of the electromagnetic field interacting with a fixed classical distribution of charge.)  Alternatively, the theory could simply be called ``quantum Dirac field theory with Coulomb interactions.''  This theory is a natural object of study, but I have not seen it examined elsewhere.  One reason this theory might be avoided is that it is not a relativistic quantum field theory (because we only include instantaneous Coulomb interactions).

Let us briefly review the quantum theory of the Dirac field without any interactions.  In this theory, the dynamics \eqref{schrodinger} are given by the Hamiltonian operator
\begin{equation}
\widehat{H}=\int :\widehat{\psi}^\dagger\big(-i \hbar c\, \gamma^0\vec{\gamma}\cdot\vec{\nabla} + \gamma^0 m c^2\big)\widehat{\psi}: d^3 x
\ .
\label{Dhamiltonian}
\end{equation}
The time-independent (Schr\"{o}dinger picture) field operator can be written in its standard plane wave expansion by putting the appropriate hats on \eqref{planewaveexpansion},\footnote{See, e.g., \cite[sec.\ 8b]{schweber1961}; \cite[pg.\ 90]{bjorkendrellfields}; \cite[pg.\ 58]{peskinschroeder}; \cite[ch.\ 5]{tong}.}
\begin{align}
\widehat{\psi} (\vec{x}) &= \overbrace{\frac{1}{(2\pi\hbar)^{3/2}}\int{ \frac{d^3 p}{\sqrt{2 \mathcal{E}(\vec{p}\,)}} \sum_{s=1}^2 \left(\widehat{b}^s (\vec{p}\,) \: u^s (\vec{p}\,) \, e^{\frac{i}{\hbar} \vec{p} \cdot \vec{x}}\right)}}^{\mbox{$\widehat{\psi}_+(\vec{x})$}}
\nonumber
\\
&\qquad+\underbrace{\frac{1}{(2\pi\hbar)^{3/2}}\int{ \frac{d^3 p}{\sqrt{2 \mathcal{E}(\vec{p}\,)}} \sum_{s=1}^2 \left(\widehat{d}^{s\dagger} (\vec{p}\,) \: v^s (\vec{p}\,) \,  e^{-\frac{i}{\hbar} \vec{p} \cdot \vec{x}}\right)}}_{\mbox{$\widehat{\psi}_-(\vec{x})$}}
\ ,
\label{operatorplanewaveexpansion}
\end{align}
where $\widehat{b}^s (\vec{p}\,)$ is an electron annihilation operator, $\widehat{d}^{s\dagger} (\vec{p}\,)$ is a positron creation operator, $\widehat{\psi}_+(\vec{x})$ is a field operator for the electron modes, and $\widehat{\psi}_-(\vec{x})$ is a field operator for the positron modes.  A single-electron quantum state can be written as a superposition of momentum eigenstates $|\vec{p}, s \rangle=\sqrt{2 \mathcal{E}(\vec{p}\,)}\;\widehat{b}^{s\dagger}(\vec{p}\,)| 0 \rangle$, taking the general form
\begin{equation}
\int d^3 p \sum_{s=1}^2 f(\vec{p},s) \sqrt{2 \mathcal{E}(\vec{p}\,)} \; \widehat{b}^{s\dagger}(\vec{p}\,)| 0 \rangle
\ ,
\label{singleelectronstate}
\end{equation}
where $| 0 \rangle$ is the zero-particle vacuum state and $f(\vec{p},s)$ is an appropriately normalized function of the momentum $\vec{p}$ and spin index $s$, determining the particular single-electron state.

We can now consider how the dynamics change if we modify the Hamiltonian \eqref{Dhamiltonian} by adding the Coulomb interaction term in \eqref{coulombnormal},
\begin{equation}
\widehat{H}=\int :\widehat{\psi}^\dagger\big(-i \hbar c\, \gamma^0\vec{\gamma}\cdot\vec{\nabla} + \gamma^0 m c^2\big)\widehat{\psi}: d^3 x+\frac{e^2}{2} \int \frac{:\widehat{\psi}^{\dagger}(\vec{x})\widehat{\psi}(\vec{x})\widehat{\psi}^{\dagger}(\vec{y})\widehat{\psi}(\vec{y}):}{|\vec{x}-\vec{y}\,|}d^3 x\,  d^3 y
\ .
\label{Dhamiltonian2}
\end{equation}
Expanding the field operators in terms of $\widehat{\psi}_+(\vec{x})$ and $\widehat{\psi}_-(\vec{x})$ from \eqref{operatorplanewaveexpansion}, the Coulomb term can be written as a sum of terms involving 0, 1, 2, 3, or 4 creation operators.  There are terms describing the creation of four particles (two electrons and two positrons), the annihilation of four particles, an electron creating an electron-positron pair, etc.  Here are the terms that leave the total numbers of electrons and positrons unchanged:
\begin{align}
&-\frac{e^2}{2} \int \sum_{i,j=1}^4\frac{\widehat{\psi}^{\dagger}_{+i}(\vec{x})\widehat{\psi}^{\dagger}_{+j}(\vec{y})\widehat{\psi}_{+i}(\vec{x})\widehat{\psi}_{+j}(\vec{y})}{|\vec{x}-\vec{y}\,|}d^3 x\,  d^3 y
\nonumber
\\
&+e^2 \int \sum_{i,j=1}^4\frac{\widehat{\psi}_{-i}(\vec{x})\widehat{\psi}^{\dagger}_{+j}(\vec{y})\widehat{\psi}^{\dagger}_{-i}(\vec{x})\widehat{\psi}_{+j}(\vec{y})}{|\vec{x}-\vec{y}\,|}d^3 x\,  d^3 y
\nonumber
\\
&-\frac{e^2}{2} \int \sum_{i,j=1}^4\frac{\widehat{\psi}_{-i}(\vec{x})\widehat{\psi}_{-j}(\vec{y})\widehat{\psi}^{\dagger}_{-i}(\vec{x})\widehat{\psi}^{\dagger}_{-j}(\vec{y})}{|\vec{x}-\vec{y}\,|}d^3 x\,  d^3 y
\ .
\label{coulombpieces}
\end{align}
The sums over the Dirac field indices $i$ and $j$, that are implicit elsewhere, need to be written out explicitly here because the normal ordering does not always place daggered operators immediately to the left of their undaggered partners.    The first term describes Coulomb repulsion between electrons, the second describes Coulomb attraction between electrons and positrons, and the last describes Coulomb repulsion between positrons.

As we are asking about the fate of electron self-repulsion and not about the interactions between electrons and positrons or among positrons, let us now focus on the one term in the expansion of the original Coulomb term \eqref{coulombnormal} that includes only \emph{electron} creation and annihilation operators,
\begin{equation}
-\frac{e^2}{2} \int \sum_{i,j=1}^4\frac{\widehat{\psi}^{\dagger}_{+i}(\vec{x})\widehat{\psi}^{\dagger}_{+j}(\vec{y})\widehat{\psi}_{+i}(\vec{x})\widehat{\psi}_{+j}(\vec{y})}{|\vec{x}-\vec{y}\,|}d^3 x\,  d^3 y
\ ,
\label{coulombelectron}
\end{equation}
the first term in \eqref{coulombpieces}.  Looking back to \eqref{operatorplanewaveexpansion}, we can recognize this term as having two electron creation operators to the left of two electron annihilation operators.  This term resembles the energy of self-repulsion for a classical cloud of electron charge in the Dirac field \eqref{electronrepulsion}, but we will see shortly that it describes only Coulomb repulsion between electrons and not electron self-repulsion.

If we time evolve a two-electron state by the Schr\"{o}dinger equation \eqref{schrodinger}, \eqref{coulombelectron} will alter its free dynamics to include electron-electron repulsion.   In the Schr\"odinger evolution of a single-electron state \eqref{singleelectronstate}, \eqref{coulombelectron} will have no effect because the sequence of two electron annihilation operators will return zero when acting on this single-particle state.  Thus, we have included electron-electron repulsion and excluded electron self-repulsion.  There is no self-repulsion in the dynamics and no contribution to the energy from self-repulsion.

If we had used the incorrect normal-ordering in \eqref{coulombnormalbad}, then the piece of the Coulomb term that contains only electron creation and annihilation operators would be
\begin{equation}
\frac{e^2}{2} \int \frac{\widehat{\psi}^{\dagger}_{+}(\vec{x})\widehat{\psi}_{+}(\vec{x})\widehat{\psi}^{\dagger}_{+}(\vec{y})\widehat{\psi}_{+}(\vec{y})}{|\vec{x}-\vec{y}\,|}d^3 x\, d^3 y
\ ,
\label{badelectronrepulsion}
\end{equation}
instead of the similar expression in \eqref{coulombelectron}.  Because \eqref{badelectronrepulsion} alternates creation and annihilation operators, it would alter the evolution of a single-electron state.  Thus, \eqref{badelectronrepulsion} does not merely describe electron-electron repulsion.  This term also includes a kind of electron self-interaction that could be called ``electron self-repulsion'' because \eqref{badelectronrepulsion} is the operator version of the classical energy of self-repulsion \eqref{electronrepulsion}.  By positing that \eqref{coulombnormal} is the correct Coulomb term to include in the Hamiltonian, not \eqref{coulombnormalbad}, we can avoid this kind of electron self-interaction.

To fully understand the kind of electron self-interaction that has been removed, and whether it indeed acts as a form of self-repulsion, would require further study.  One question that could be asked is whether a lone electron would rapidly expand as its parts repel one another.  That question can be put more precisely.  One way of understanding the quantum state that evolves by the Schr\"{o}dinger equation \eqref{schrodinger} is as a wave functional that describes a quantum superposition of different classical field configurations \cite{fields}.  Each classical configuration of the Dirac field gives a precise charge distribution and thus a superposition of field configurations is also a superposition of different charge distributions.  One could begin by using a superposition of compact charge distributions to represent a lone electron in empty space and then ask whether this superposition would evolve by the Schr\"{o}dinger equation \eqref{schrodinger} into a superposition of more widely-spread wave packets, seeing if the above electron self-interaction \eqref{badelectronrepulsion} would yield more rapid expansion than might occur under the free dynamics \eqref{Dhamiltonian}.  To better understand the above self-interaction, one could also ask how it would change the shapes of atoms.  For example, one could study the ground state of the hydrogen atom and see if the superposed electron charge distributions are spread wider when the partially normal-ordered Coulomb term \eqref{coulombnormalbad} is used instead of the fully normal-ordered Coulomb term \eqref{coulombnormal}.  The electron cloud should be larger (and the ground state energy higher) because, with self-repulsion included, the inner part of the electron's charge distribution would partially shield the outer part from the charge of the nucleus.

Let us set that self-interaction aside and return to the theory of quantum electrostatics under investigation here (using the correct normal-ordering).  With the full Coulomb interaction term \eqref{coulombnormal} included in the Hamiltonian \eqref{Dhamiltonian2}, not just the electron-only term in \eqref{coulombelectron} or the particle-number-preserving terms in \eqref{coulombpieces}, the zero-particle vacuum state $| 0 \rangle$ from the free theory is no longer the ground state (a common feature of quantum field theories with interactions\footnote{See the discussion of Haag's theorem in \cite[sec.\ 3]{earman2006}; \cite[sec.\ 3]{fraser2008fate}; \cite[sec.\ 4.3]{fields}.}).  The Coulomb term will generate particles from the vacuum.  There should be a minimum energy ground state $| \Omega \rangle$ that evolves trivially under the dynamics \eqref{schrodinger} (only changing its global phase), with $\widehat{H}| \Omega \rangle = E_0 | \Omega \rangle$, where $E_0$ is the ground state energy.  Depending on what we deem worthy of the title ``particle,'' one might either say that this ground state of the interacting theory is the zero-particle vacuum state for that theory (even though this state is distinct from the zero-particle state of the free theory) or that the ground state of the interacting theory has particle content that can be treated as a background upon which we can look for deviations.  For our purposes here, let us adopt the latter interpretation and view the ground state as containing particles.  On this interpretation, one can ask about the evolution of what might be called ``single-extra-electron states'' that are arrived at by acting on the minimum energy ground state $| \Omega \rangle$ with electron creation operators as in \eqref{singleelectronstate}.  These are states where an electron has been added to the particle content of the ground state.  The evolution for these states will be more complicated than for the simple single-electron states discussed above.  The electron-only term in \eqref{coulombelectron} will continue to omit self-repulsion, but it will include Coulomb repulsion between the extra electron and background electrons.
 
Moving from the quantum theory of the Dirac field with Coulomb interactions that we have been discussing to full quantum electrodynamics, quantum states become more complicated (as they describe both the Dirac and electromagnetic fields) and the evolution of states becomes more complicated (as the Hamiltonian includes more terms).  The Hamiltonian of quantum electrodynamics in the Coulomb gauge can be written as
\begin{align}
H&=\int \left(\frac{:\big|\frac{\partial\hat{\vec{A}}}{\partial t}\big|^2:}{8 \pi c^2}
 + \frac{:\big(\vec{\nabla} \times \hat{\vec{A}}\,\big)^2:}{8 \pi} +:\widehat{\psi}^\dagger\big(-i \hbar c\, \gamma^0\vec{\gamma}\cdot\vec{\nabla} + \gamma^0 m c^2\big)\widehat{\psi}:\right.
 \nonumber
 \\
&\qquad\quad\left. + :e\, \widehat{\psi}^\dagger\gamma^0\vec{\gamma}\widehat{\psi} \cdot \hat{\vec{A}}:+ \frac{e^2}{2} \int \frac{:\widehat{\psi}^{\dagger}(\vec{x})\widehat{\psi}(\vec{x})\widehat{\psi}^{\dagger}(\vec{y})\widehat{\psi}(\vec{y}):}{|\vec{x}-\vec{y}|} d^3 y\right)d^3 x
\ ,
\label{QEDhamiltonian}
\end{align}
putting hats on the expression for the energy of the interacting Dirac and Maxwell fields in \eqref{hamiltonian2} and normal-ordering every term.  This Hamiltonian includes the Coulomb interaction term \eqref{coulombnormal} that we have been analyzing.  In this theory, the evolution for a single-electron state \eqref{singleelectronstate} will be altered from the free evolution by both the Coulomb term---though not the parts in \eqref{coulombpieces}---and the $:e\, \widehat{\psi}^\dagger\gamma^0\vec{\gamma}\widehat{\psi} \cdot \widehat{\vec{A}}:$ term (which should account for radiation reaction).

\section{Conclusion}

If the electron is modeled classically as a cloud of energy and charge in the Dirac field, then it will experience self-repulsion (section \ref{CFTsection}).  This self-repulsion can be eliminated in quantum field theory by fully normal-ordering the Coulomb term in the Hamiltonian operator (section \ref{QFTsection}).  The Hamiltonian of quantum field theory includes Coulomb attraction and repulsion between distinct particles and excludes self-repulsion.

\vspace*{12 pt}
\noindent
\textbf{Acknowledgments}
Thank you to Jacob Barandes, Maaneli Derakhshani, Michael Miller, Logan McCarty, Simon Streib, and anonymous reviewers for helpful feedback and discussion.

\end{document}